\documentclass[aps,prb,twocolumn,amsmath,amssymb,floatfix]{revtex4}

\usepackage[colorlinks,citecolor=blue,linkcolor=blue,urlcolor=blue,plainpages=false,pdfpagelabels]{hyperref}
\usepackage{txfonts}

\begin{document}

\title{Entropy production in one-dimensional quantum fluids}

\author{Edvin G. Idrisov}
\author{Thomas L. Schmidt}

\affiliation{Physics and Materials Science Research Unit, University of Luxembourg, L-1511 Luxembourg, Luxembourg}

\date{\today}

\begin{abstract}

We study nonequilibrium thermodynamic properties of a driven one-dimensional quantum fluid by combining nonlinear Luttinger liquid theory with the quantum kinetic equation. In particular, we derive an entropy production consistent with the laws of thermodynamics for a system subject to an arbitrary perturbation varying slowly in space and time. Working in a basis of weakly interacting fermionic quasiparticles, we show that the leading contribution to the entropy production results from three-particle collisions, and we derive its scaling law at low temperatures.

\end{abstract}

\maketitle

\section{Introduction}
\label{Sec:I}

One-dimensional (1D) fermionic systems play an important role in modern condensed-matter physics because they display phenomena which are starkly different from those seen in higher dimensional systems.\cite{giamarchi,delft98} Electron-electron interactions have a strong impact on 1D systems because the restricted dimensionality enhances scattering, and they ultimately destroy the simple quasiparticle picture of Landau's Fermi liquid theory which has been very successful in higher dimensions. The conducting state of a 1D quantum system is called a Luttinger liquid (LL)\cite{haldane81} and recent developments in the experimental fabrication of electronic 1D systems, for instance in carbon nanotubes,\cite{bockrath99,shi15} semiconductor nanowires,\cite{auslaender05,jompol09} or quantum Hall edge states\cite{ezawa} have driven experimental and theoretical investigation in this field.

The bosonization technique\cite{mattis65} is a powerful tool for the theoretical description of interacting fermionic 1D systems in the low-energy sector.\cite{tomonaga50,luttinger63,haldane81} Its starting point is to split the physical fermion field $\Psi(x)$ in the vicinity of the Fermi points $\pm k_F$ into chiral right-moving ($R$) and left-moving ($L$) fermion fields $\Psi_{R,L}(x)$, and to express these operators in terms of density-like collective bosonic excitations, $\phi(x)$ and $\theta(x)$, as $\Psi_{R,L}(x) \propto \exp(-i[\pm \phi(x)-\theta(x)])$. The bosonic fields satisfy the canonical commutation relation $[\phi(x),\partial_{x^{\prime}} \theta(x^{\prime})]=i\pi \delta(x-x^{\prime})$. Linearizing the spectrum of the right-movers and left-movers near the Fermi points, $\epsilon_{R,L}(k) \approx v_F( \pm k - k_F)$ where $v_F$ is the Fermi velocity, the total Hamiltonian of the 1D system $H_{LL}=H_{kin}+H_{int}$, consisting of kinetic energy and interaction energy, takes a quadratic form in boson fields
\begin{equation}
\label{Total Hamiltonian in terms of bosons with linear spectrum}
H_{LL}=\frac{v}{2\pi} \int dx \left[K(\partial_x \theta)^2+\frac{1}{K}(\partial_x \phi)^2\right].
\end{equation}
In this Luttinger liquid Hamiltonian, $v$ is a Fermi velocity renormalized by the interactions, and $K$ is
the Luttinger parameter, which for fermions with repulsive interactions is between zero and one. For the noninteracting system, one finds $K=1$ and $v=v_F$. The bosonization approach thus allows an exact diagonalization of the Hamiltonian of interacting fermions with linear spectrum in terms of bosonic fields, and even makes it possible to calculate dynamic correlation functions.

Luttinger theory has been very successful in describing zero-energy properties of 1D systems. However, to explain phenomena for which finite-energy excitations are important, such as relaxation and equilibration in quantum nanowires,\cite{karzig,matveev10,micklitz10,levchenko11,micklitz11,matveev12,matveev12b,matveev13a,
ristivojevic13,protopopov14,matveev17} Coulomb drag between quantum wires,\cite{dmitriev12} or momentum-resolved tunneling of electrons in nanowires,\cite{imambekov09,schmidt09_3,schmidt10_2,imambekov12} one has to go beyond the approximation of linear spectrum, and needs to take into account its curvature, which is typically quadratic near the Fermi points,
\begin{align}\label{spectrum_quadratic}
    \epsilon_{R,L}(k) \approx v_F(\pm k - k_F) + \frac{1}{2m^*} \left(\pm k-k_F \right)^2,
\end{align}
with an effective mass $m^*$. After bosonization, the spectrum curvature is found to induce interactions between the aforementioned bosonic modes. Various methods have been proposed to tackle problems with nonlinear dispersion relation $\epsilon_{R,L}(k)$, which are subsumed under the name of nonlinear Luttinger liquid (NLL) theory.\cite{imambekov12}

One of those techniques is refermionization.\cite{imambekov12,rozhkov05,rozhkov06,imambekov09} It allows one to map the physical fermions with spectrum curvature and arbitrary interaction strength onto \emph{weakly interacting} fermionic quasiparticles.\cite{rozhkov05,rozhkov06} In contrast to the interactions between the physical fermions, the interactions between the quasiparticles are irrelevant in the renormalization group (RG) sense, so one can then apply the conventional perturbation theory to calculate observables such as for instance response functions.\cite{rozhkov05,imambekov12} In the limit of a strictly linear spectrum, the fermionic quasiparticles become noninteracting.

The effects of relaxation processes on electron transport in 1D systems with nonlinear spectrum were studied in Refs.~[\onlinecite{karzig,matveev10,micklitz10,levchenko11,micklitz11,matveev12,matveev12b,matveev13a,
ristivojevic13,protopopov14,matveev17}]. Spectrum curvature is essential for relaxation, but it was already pointed out in these works that for the most realistic case of a parabolic spectrum as in Eq.~(\ref{spectrum_quadratic}), kinematic constraints forbid relaxation due to two-particle collisions. Hence, one generally needs to take into account at least three-particle collisions, and by studying those it was demonstrated that the decay rate of fermionic quasiparticles in 1D is drastically different from the predictions of Landau's Fermi liquid theory.\cite{matveev13a}

More recently, attention has shifted towards a study of thermal transport in 1D electronic fluids using both bosonic and fermionic approaches.\cite{matveev19,samanta19} The steady appearance of new results indicates that the physics of 1D quantum systems is not yet fully understood. Our work considers driven, nonequilibrium thermodynamic properties of interacting 1D systems with nonlinear spectrum. Thermodynamic behavior is to a large extent governed by changes of the entropy of the system. This is why the investigation of the entropy production in a 1D quantum system has been chosen to be the main subject of this article.

In order to theoretically study thermodynamic properties like entropy flow and entropy production, we combine refermionization with a real-time nonequilibrium Green's function approach. To take into account the interaction between quasiparticles and an external space- and time-dependent potential we use the gradient approximation\cite{rammer,kita} and derive kinetic equations for the system under consideration. This allows us to provide an expression for the entropy production which satisfies the second law of thermodynamics.

The gradient approximation allows a systematic expansion in the rate of change of the external perturbation in space and time. Its zeroth order corresponds to an adiabatic evolution, whereas the first-order gradient approximation we will use corresponds to the leading non-adiabatic correction for slow driving. Hence, we would like to point out that the limit we are considering is opposite to that of a quench, where a sudden change of an external potential is assumed.\cite{Cazalilla06,Calabrese06,iucci09,Essler16,calzona17,calzona17a,calzona18}

Besides the gradient approximation, our approach rests on perturbation theory in the residual weak interactions between the refermionized quasiparticles. In contrast to the interactions between the physical fermions, those allow a perturbative treatment, and it was shown before that such an approach is applicable for energies much less than the Fermi energy.\cite{protopopov14} For larger energies, on the other hand, perturbation theory in the bosonic basis would be more appropriate. We would like to point out that our approach is valid for arbitrary interaction strength between the \emph{physical} fermions.

The rest of the paper is organized as follows. In Sec.~\ref{Sec:Model}, we define the model we consider. In Sec.~\ref{Sec:KineticEquation}, we derive the exact quantum kinetic equation for the nonlinear Luttinger liquid. In Sec.~\ref{Sec:Entropy}, we use it to derive the continuity equation for the entropy density and define the entropy production. We present our conclusions in Sec.~\ref{Sec:Conclusion}. Details of the calculations are presented in appendices. Throughout the paper, we set $e=\hbar=k_B=1$.

\section{Theoretical model}
\label{Sec:Model}
\subsection{Hamiltonian}
\label{Hamiltonian}
We consider an interacting 1D system of spinless fermions which are subject to an external perturbation which couples to the particle density. The total Hamiltonian of this system is given by
\begin{equation}
\label{Total physical Hamiltonian}
H(t)=H_{kin}+H_{int}+H_{ext}(t).
\end{equation}
The kinetic energy term reads
\begin{equation}
\label{Physical kinetic Hamiltonian}
H_{kin}=\int dx \Psi^{\dagger}(x)\left(\frac{\hat{p}^2}{2m}-\mu\right)\Psi(x),
\end{equation}
where $\Psi^{\dagger}(x)$ and $\Psi(x)$ are creation and annihilation operators for physical fermions at position $x$, satisfying the anticommutation relations $\{\Psi^{\dagger}(x),\Psi(y)\}=\delta(x-y)$ and $\{ \Psi(x),\Psi(y)\}=0$. Moreover, $\hat{p} =-i\partial_x$ is the momentum operator in 1D, $m$ is the fermion mass, and $\mu$ is the chemical potential. The second term in Eq.~(\ref{Total physical Hamiltonian}) represents the repulsive interactions between the physical fermions. It is a functional of the density and has the form
\begin{equation}
\label{Physical interaction Hamiltonian}
H_{int}=\frac{1}{2} \int dx \int dx^{\prime} \Psi^{\dagger}(x)\Psi^{\dagger}(x') V(x-x^{\prime}) \Psi(x')\Psi(x),
\end{equation}
where $V(x-x^{\prime})$ is a generic two-body interaction potential. We assume that the Fourier transform $V(k)$ of this potential at $k=0$ is finite, thus ruling out unscreened Coulomb interactions. The last term in the full Hamiltonian~(\ref{Total physical Hamiltonian}),
\begin{equation}\label{eq:Hext}
H_{ext}(t)=\int dx U(x,t) \Psi^{\dagger}(x)\Psi(x),
\end{equation}
describes the effect of an applied external field which is coupled to charge density. Expressing the time-independent Hamiltonian $H_{kin} + H_{int}$ in terms of right- and left-movers,
\begin{align}
    \Psi(x) = e^{i k_F x} \Psi_R(x) + e^{-i k_F x} \Psi_L(x),
\end{align}
approximating the kinetic energy to linear order in momentum around $\pm k_F$, and applying the bosonization formula, one arrives at the Luttinger Hamiltonian~(\ref{Total Hamiltonian in terms of bosons with linear spectrum}), which is exactly solvable.

\subsection{Nonlinear Luttinger liquid}
\label{nonlinear Luttinger liquid}

The Luttinger Hamiltonian (\ref{Total Hamiltonian in terms of bosons with linear spectrum}) is diagonal in terms of bosonic eigenmodes and thus integrable. As the Hamiltonian corresponding to the external potential~(\ref{eq:Hext}) is diagonal in these eigenmodes as well, each Fourier component of the external potential $U(x,t)$ affects only the individual bosonic mode with the corresponding wave vector and frequency. Hence, within linear Luttinger liquid theory, one obtains a collection of uncoupled, individually driven bosonic modes. Such a system lacks relaxation and its entropy will be constant. A nonzero entropy production in our closed system requires interactions between the different modes, caused for instance by spectrum curvature.

The exact kinetic energy (\ref{Physical kinetic Hamiltonian}) of the physical fermions is not a linear function of momentum. To account for its curvature, $H_{LL}$ has to be supplemented with correction terms. It can be shown that these corrections are RG-irrelevant, so the Luttinger liquid picture at low energies is in principle justified.\cite{haldane81} Nonetheless, it is known that these corrections play an important role for relaxation processes because they permit the decay of the collective bosonic excitations.\cite{matveev13a} If one attempts to translate fermionic curvature terms into the bosonic language, one recovers cubic terms in the fields $\phi$ and $\theta$, rendering the bosonic theory interacting. Unfortunately, perturbation theory in the bosonic theory leads to divergences and the partial resummation of diagrams is generally a difficult task.\cite{teber07}

Instead of using the bosonic language, it is often more convenient to develop a theory which is based on \emph{fermionic} quasiparticles.\cite{rozhkov05,rozhkov06,imambekov09,imambekov12} In the noninteracting limit, they coincide with the physical fermions, but in the interacting case, they are related to them via a nonlocal unitary transformation.\cite{rozhkov05} This direct refermionization procedure allows us to rewrite the total Hamiltonian~(\ref{Total physical Hamiltonian}) at low energies in terms of new right-moving and left-moving fermion quasiparticles. The total Hamiltonian then includes two terms
\begin{equation}
\label{Total Hamiltonian of fermion quasiparticles}
H(t)=\mathcal{H}_0(t)+\mathcal{H}_{int},
\end{equation}
where $\mathcal{H}_0(t)$ denotes the noninteracting quasiparticle Hamiltonian including the external perturbation,
\begin{equation}
\label{noninteracting Hamiltonian}
\begin{split}
& \mathcal{H}_0(t)= \sum_{\alpha=L,R} \int dx \psi^{\dagger}_{\alpha}(x) \mathcal{K}_{\alpha}(x,t)\psi_{\alpha}(x), \\
& \mathcal{K}_{\alpha}(x,t)=-\frac{\partial_x^2}{2 \tilde{m}}-i\alpha \tilde{v}_F \partial_x +\sqrt{K}U(x,t),\\
\end{split}
\end{equation}
where $\tilde{v}_F$ is a renormalized Fermi velocity and $\tilde{m}$ is an effective quasiparticle mass, where $\tilde{m}=4m/(K\sqrt{K}+3/\sqrt{K})$ for weak interactions. \cite{imambekov12} The fermion quasiparticle operators satisfy the conventional anticommutation relations $\{\psi^{\dagger}_{\alpha}(x),\psi_{\beta}(y)\}=\delta_{\alpha \beta}\delta(x-y)$ and $ \{ \psi_{\alpha}(x),\psi_{\beta}(y)\}=0$.

In addition, the curvature of the spectrum of the physical fermions leads to a two-body interaction term between quasiparticles on opposite branches
\begin{equation}
\label{Interaction Hamiltonian of fermion quasiparticles}
\mathcal{H}_{int}=i\tilde{g} \sum_{\alpha=\pm} \alpha \int dx \rho_{-\alpha}\left \{ \psi^{\dagger}_{\alpha}(\partial_x \psi_{\alpha})-(\partial_x \psi^{\dagger}_{\alpha})\psi_{\alpha} \right\},
\end{equation}
where $\rho_{\alpha}=\psi^{\dagger}_{\alpha}\psi_{\alpha}$ is the quasiparticle density (for $\alpha = R, L = +, -$). The strength of the interaction is given by  $\tilde{g}=\pi(K^{3/2}-K^{-1/2})/4m$. Moreover, refermionization also reveals an interaction term between quasiparticles on the same branch, but the latter has a higher scaling dimension than $\mathcal{H}_{int}$ and can therefore be neglected at low energies.\cite{imambekov09} It is worth pointing out that the mapping between interacting physical fermions and fermionic quasiparticles can also be performed using the bosonization procedure as an intermediate step and produces the same result.\cite{imambekov12}

For further consideration it is convenient to rewrite Eq.~(\ref{Interaction Hamiltonian of fermion quasiparticles}) in momentum space,
\begin{equation}
\label{Interaction Hamiltonian of fermion quasiparticles in momentum representation}
\mathcal{H}_{int}=\frac{\tilde{g}}{L}\sum_{\substack{kk^{\prime} \\ pp^{\prime}}} (p+p^{\prime}-k-k^{\prime})\delta_{p+k,p^{\prime}+k^{\prime}}c^{\dagger}_{Lk}c_{Lk^{\prime}}c^{\dagger}_{Rp}c_{Rp^{\prime}},
\end{equation}
where the Kronecker delta is a consequence of momentum conservation and $L$ in the prefactor is the system length. Equation~(\ref{Interaction Hamiltonian of fermion quasiparticles in momentum representation}) describes an effective two-body interaction with scattering amplitude linear in the momentum. This entails that $\mathcal{H}_{int}$ corresponds to an RG-irrelevant interaction and is thus amenable to perturbation theory.

The Hamiltonian (\ref{Total Hamiltonian of fermion quasiparticles}) with interaction term given by Eq.~(\ref{Interaction Hamiltonian of fermion quasiparticles in momentum representation}) gives the complete description of our system. For $U(x,t) = 0$, this Hamiltonian has been used to study the decay of fermionic quasiparticles in 1D quantum liquids.\cite{matveev13a}

\section{Quantum kinetic equation}
\label{Sec:KineticEquation}

\subsection{Wigner transformation}
\label{Wigner transformation}

We now introduce the necessary ingredients to derive a quantum kinetic equation for 1D electron fluids. The essential approximation allowing us to proceed analytically is to assume that the external perturbation $U(x,t)$ varies slowly. The system degrees of freedom can then be split into rapid and slow modes,\cite{rammer,kita,ivanov} which suggests to use the Wigner representation of the Keldysh Green's function matrix,
\begin{equation}
\check{G}_{\alpha}(k\varepsilon, xt) \equiv \int d \delta x \int d \delta t \check{G}_{\alpha}(1,2)e^{-i(k \delta x-\varepsilon \delta t)},
\end{equation}
where we used the notation $1\equiv(x_1,t_1)$ and we have introduced the center-of-mass $x=(x_1+x_2)/2$, $t=(t_1+t_2)/2$ and relative $\delta x=x_1-x_2$, $\delta t=t_1-t_2$ coordinates. The Keldysh Green's function $\check{G}_{\alpha}(1,2)$ is defined in Eq.~(\ref{Keldysh matrix}) in App.~\ref{Sec:A}. The Wigner transformation is nothing but a partial Fourier transformation with respect to the relative coordinates. Next, we define the spectral functions $A_{R,L}(k\varepsilon,xt)$ and the quasiparticle distribution functions $\phi_{R,L}(k\varepsilon,xt)$ by
\begin{equation}
\label{Definition of spectral and distribution functions}
\begin{split}
& G^{-+}_{\alpha}(k\varepsilon,xt)=iA_{\alpha}(k\varepsilon,xt)\phi_{\alpha}(k\varepsilon,xt), \\
& G^{+-}_{\alpha}(k\varepsilon,xt)=-iA_{\alpha}(k\varepsilon,xt)[1-\phi_{\alpha}(k\varepsilon,xt)].
\end{split}
\end{equation}
From the symmetry of spectral function $A^{\ast}_\alpha(1,2)=A_\alpha(2,1)$, and the commutation relations of the fermion quasiparticles at equal times it follows that $A^{\ast}_{\alpha}(k\varepsilon,xt)=A_{\alpha}(k\varepsilon,xt)$ and that it is normalized as $\int d\varepsilon A_{\alpha}(k\varepsilon,xt)=2\pi$. Moreover, one can show that $A_{\alpha}(k\varepsilon,xt) \geq 0$ and that $0 \leq \phi_{\alpha}(k\varepsilon,xt) \leq 1$.\cite{kita} These two quantities can be considered as an alternative pair of functions to the off-diagonal elements $G^{-+}_{\alpha}$ and $G^{+-}_{\alpha}$.

\subsection{Gradient approximation}
To obtain the kinetic equation one needs to apply the Wigner transformation to differential form of Dyson's equation [see Eq.~(\ref{Differential and integral Dyson equations}) in App.~\ref{Sec:A}]. It is well known that convolutions are transformed as follows by the Wigner transformation,\cite{rammer,kita,ivanov,arseev}
\begin{equation}
\int C(1,3) D(3,2) d3 = \int\frac{dk d\varepsilon}{(2\pi)^2} [C \ast D] e^{i(k \delta x-\varepsilon \delta t)},
\end{equation}
where on the right hand side $C=C(k\epsilon,xt)$ and $D=D(k\epsilon,xt)$. The asterisk operator on the right hand side denotes a Moyal product and is defined as
\begin{equation}
C \ast D \equiv C\exp\left[\frac{i\hat{O}}{2}\right]D,
\end{equation}
where the differential operator in the exponent is given by
\begin{equation}
\hat{O}=\overleftarrow{\partial_x}\ \overrightarrow{\partial_k}-\overleftarrow{\partial_t}\ \overrightarrow{\partial_{\varepsilon}}-\overleftarrow{\partial_k}\  \overrightarrow{\partial_x}+\overleftarrow{\partial_{\varepsilon}}\ \overrightarrow{\partial_t}.
\end{equation}
The left (right) arrow on each differential operator denotes that it acts towards the left (right) side of the expression. An exact calculation of the Moyal product is not possible, but it allows for a systematic expansion in orders of temporal and spatial derivatives. Performing a Taylor expansion with respect to the center-of-mass coordinates and keeping only the first term corresponds to the first-order gradient expansion,
\begin{equation}
\label{Gradient approximation}
C \ast D \approx C D+\frac{i}{2} \left\{C,D\right\},
\end{equation}
where braces correspond to Poisson brackets, namely
\begin{equation}
\label{Poisson brackets}
\left\{C,D\right\} \equiv \frac{\partial C}{\partial x}\frac{\partial D}{\partial k}-\frac{\partial C}{\partial t}\frac{\partial D}{\partial \varepsilon}-\frac{\partial C}{\partial k}\frac{\partial D}{\partial x}+\frac{\partial C}{\partial \varepsilon}\frac{\partial D}{\partial t}.
\end{equation}
The gradient approximation holds when the external perturbation $U(x,t) \propto \int d\omega \int dq e^{i(qx-\omega t)}U(q,\omega)$ varies slowly in space and time, so that its characteristic frequency $\omega \ll \tilde{\varepsilon}_F=\tilde{m}\tilde{v}^2_F/2$ and wave vector $q \ll \tilde{k}_F=\tilde{m}\tilde{v}_F$.

\subsection{Transport equation}
\label{Transport equation}
Now we are ready to derive the quantum kinetic equation from Dyson's equation. To do this we rewrite the Dyson equation~(\ref{Differential and integral Dyson equations}) in Wigner representation using the Moyal product
\begin{equation}
\label{Zero order Green;s function in Wigner representation}
\left(G^{-1}_{0,\alpha} \check{I}-\check{\Sigma}^K_{\alpha}\right) \ast \check{G}^K_{\alpha}=\check{I},
\end{equation}
where $\check{I}$ is a $2\times 2$ identity matrix and superscripts $K$ indicate Green's function matrices in the Larkin-Ovchinnikov basis [see Eq.~(\ref{Keldysh matrix after rotation})],\cite{larkin}
\begin{align}
    \check{G}^K_{\alpha}(1,2) = \begin{pmatrix}
       G^{R}_{\alpha} & G^{K}_{\alpha}         \\[0.3em]
       0 & G^{A}_{\alpha}           \\[0.3em]
     \end{pmatrix},
\end{align}
consisting of retarded (R), advanced (A), and Keldysh (K) components. The self-energy $\check{\Sigma}^K_\alpha$ has an analogous structure. The inverse of the unperturbed Green's function has the form
\begin{equation}
G^{-1}_{0,\alpha}(k\varepsilon,xt)=\varepsilon - \xi_{k,\alpha},
\end{equation}
where $\xi_{k,\alpha}=k^2/2\tilde{m}+\alpha \tilde{v}_F k+\sqrt{K}U(x,t)$. Using the gradient expansion~(\ref{Gradient approximation}) and the definition of spectral and distribution functions~(\ref{Definition of spectral and distribution functions}), we obtain after some algebra the quantum kinetic equation
\begin{equation}
\label{Quantum kinetic equation}
\{G^{-1}_{0,\alpha}-\text{Re}\Sigma^R_{\alpha}, A_{\alpha}\phi_{\alpha}\}+i\{\Sigma^{-+}_{\alpha}, \text{Re}G^{R}_{\alpha}\}=\mathcal{I}_{\alpha},
\end{equation}
with the collision integral,
\begin{equation}
\label{Scattering integral for right and left movers}
\mathcal{I}_{\alpha}=G^{+-}_{\alpha} \Sigma^{-+}_{\alpha}-G^{-+}_{\alpha}\Sigma^{+-}_{\alpha}.
\end{equation}
For given $A_{\alpha}$ and self energy matrix $\check{\Sigma}_{\alpha}$, Eq.~(\ref{Quantum kinetic equation}) gives the exact solution for the distribution function $\phi_{\alpha}$. Note that right- and left-movers are coupled by the self-energy in the collision integral. Relying only on Dyson's equation, the quantum kinetic equation is exact in the interactions, but in an interacting system it is generally not exactly solvable. However, it can be used to derive conservation laws. Hence, in the next section, the quantum kinetic equation will be used to derive a continuity equation for the entropy density.

\section{Entropy production}
\label{Sec:Entropy}

The entropy production is one of the central quantities in nonequilibrium thermodynamics and is at the origin of the irreversibility of thermodynamic processes. The first law of thermodynamics reflects energy conservation and relates the change in internal energy $\dot{U}$ to the work and heat flows into the system by $\dot{U} = \dot{W} + \dot{Q}$. The work rate is defined as $\dot{W} = \text{Tr}[\rho(t) (\partial_t H)]$ with the density matrix $\rho(t)$ and the full Hamiltonian $H(t)$.\cite{campisi11} For a system weakly coupled to a thermal bath, $\dot{Q}$ is related to the change of energy in the bath.\cite{esposito10}

Reversibility is governed by the second law of thermodynamics. It relates the change in system entropy $S$ to the heat $Q$ and the entropy production $\Delta S$. In differential form, it is given by $\dot{S} = \dot{Q}/T + \Delta \dot{S}$ where $T$ is the temperature. If the system is in equilibrium or if the external perturbation is adiabatic, the entropy production vanishes $\Delta \dot{S} = 0$. In the general case, thermodynamics requires that $\Delta \dot{S} \geq 0$. The challenge of statistical physics is to find microscopic definitions of the thermodynamic quantities which fulfill these laws. Here, we will do this for interacting 1D quantum systems.

A direct measurement of the entropy production is difficult, but $\Delta S$ is in fact related to more accessible quantities. Assuming the system to have constant temperature and constant volume, one introduces the free energy via $F = U - TS$. Eliminating the heat using the second law, one finds the standard relation,\cite{esposito10}
\begin{align}\label{eq:DeltaS_WF}
    \Delta \dot{S} = \dot{W} - \dot{F},
\end{align}
which directly relates the entropy production to experimentally more accessible quantities like the work input and the change in free energy.

\subsection{Continuity equation}

The quantum kinetic equation enables us to derive a continuity equation relating the entropy density and the entropy current to the entropy production.\cite{kita}

To this end, we multiply Eq.~(\ref{Quantum kinetic equation}) by the factor $\log[(1-\phi_{\alpha})/\phi_{\alpha}]$, perform an integration over the variables $k$ and $\varepsilon$, and exploit the fact that $0 = \text{Im}\{(G^R_{\alpha})^{-1},G^R_{\alpha}\}=\text{Im}\{G^{-1}_{0,\alpha}-\Sigma^R_{\alpha},G^R_{\alpha}\}$. These steps lead to the following continuity equation (see Appendix~\ref{Sec:B} for details)
\begin{equation}
\label{Continuity equation}
\frac{\partial s_{\alpha}}{\partial t}+\frac{\partial j_{\alpha}}{\partial x}  = \left[\frac{\partial \mathcal{S}_{\alpha}}{\partial t}\right]_{coll},
\end{equation}
whose individual components are given by
\begin{align}
\label{Entropy production for right and left mover}
 &\left[\frac{\partial \mathcal{S}_{\alpha}}{\partial t}\right]_{coll} \equiv \int \frac{dk d\varepsilon}{(2\pi)^2} \mathcal{I}_{\alpha}\log\left(\frac{1-\phi_{\alpha}}{\phi_{\alpha}}\right),\\
\label{Entropy density}
&s_{\alpha}= \int \frac{dk d\varepsilon}{(2\pi)^2}\sigma_{\alpha} \left[A_{\alpha}\frac{\partial B_{\alpha}}{\partial \varepsilon}+A_{\Sigma_{\alpha}}\frac{\partial \text{Re}G^R_{\alpha}}{\partial \varepsilon}\right], \\
\label{Entropy current}
&j_{\alpha} = \int \frac{dk d\varepsilon}{(2\pi)^2}\sigma_{\alpha} \left[-A_{\alpha}\frac{\partial B_{\alpha}}{\partial k}-A_{\Sigma_{\alpha}}\frac{\partial \text{Re}G^R_{\alpha}}{\partial k}\right],
\end{align}
where $A_{\Sigma_{\alpha}}=-i\phi^{-1}_{\alpha} \Sigma^{-+}_{\alpha}$, $B_{\alpha}=G^{-1}_{0,\alpha}-\text{Re}\Sigma^R_{\alpha}$ and $\sigma_\alpha[\phi_{\alpha}]$ corresponds to the Shannon entropy associated with the distribution $\phi_{\alpha}$,
\begin{equation}
\label{General expression for entropy}
\sigma_{\alpha}[\phi_{\alpha}]\equiv -\phi_{\alpha}\log \phi_{\alpha} - (1 - \phi_{\alpha})\log(1 - \phi_{\alpha}).
\end{equation}
Eqs.~(\ref{Entropy density}) and (\ref{Entropy current}) may indeed be considered as entropy density and entropy flux density. As shown in Ref.~[\onlinecite{kita}], in equilibrium, $s_\alpha$ coincides with the thermodynamic entropy obtained from the grand canonical potential $\Omega=-T \log \text{Tr }e^{-H/T}$. This is easiest to see in the limit of free particles. Assuming $\check{\Sigma}_\alpha \equiv 0$, we have
\begin{align}
    s^{(0)}_\alpha(xt) = \int \frac{dkd\varepsilon}{(2\pi)^2} A_\alpha(k\varepsilon,xt) \sigma_\alpha(k\varepsilon,xt)
\end{align}
which is of the same form as the entropy found in Refs.~[\onlinecite{bruch16,haughian18}]. In equilibrium ($U(x,t) = 0$), we then find $A_\alpha = 2\pi \delta[\varepsilon - k^2/(2m) - \alpha v_F k]$ and $\phi_\alpha$ becomes a Fermi distribution. It is then easy to see that $s_\alpha$ coincides with the von-Neumann entropy, and the total entropy coincides with that calculated by Rozhkov in Ref.~[\onlinecite{rozhkov05,rozhkov06}],
\begin{align}
\label{Entropy density in equilibrium within nonlinear spectrum}
\frac{S}{L} & =\frac{1}{L} \int dx (s_R + s_L) = \int \frac{dk}{2\pi}(\sigma_R+\sigma_L) \notag \\
&\simeq \frac{T}{3 \tilde{v}_F} +\frac{14\pi^3}{15}\frac{1}{(2\tilde{m})^2}\frac{T^3}{\tilde{v}^5_F}.
\end{align}
The first term corresponds to LL with linear spectrum~\cite{giamarchi} and the second term is a sub-leading correction due to spectrum curvature.

The right-hand side of continuity equation~(\ref{Continuity equation}) corresponds to the entropy production per unit time and unit length due to the quasiparticle collisions brought about by the interaction Hamiltonian~(\ref{Interaction Hamiltonian of fermion quasiparticles}). The total entropy production is given by the sum of right-mover and left-mover contributions
\begin{equation}
\label{Total entropy production}
\left[\frac{\partial \mathcal{S}}{\partial t}\right]_{coll}=\left[\frac{\partial \mathcal{S}_R}{\partial t}\right]_{coll} +\left[\frac{\partial \mathcal{S}_L}{\partial t}\right]_{coll}.
\end{equation}

\label{Sec:V}
To calculate the entropy production rate we need to know the form of scattering integral $\mathcal{I}_{\alpha}$ in Eq.~(\ref{Scattering integral for right and left movers}). The calculation of self-energies in Eq.~(\ref{Scattering integral for right and left movers}) can be performed
using a so-called self-consistent dressed (``skeleton'') Feynman diagram expansion.\cite{kita} At a given order in perturbation theory, ``skeleton'' diagrams correspond to a partial resummation of the perturbation series, where we only keep diagrams without the self-energy insertions in the expansion of the self-energy and replace unperturbed Green's functions $G^{ij(0)}_{\alpha}$ by the exact ones $G^{ij}_{\alpha}$. The details of this approach are provided in Refs.~[\onlinecite{luttinger60,luttinger60a,baym,kita}]. The diagonal elements of self energy matrix (see Eq.~(\ref{Self energy matrix}) in Appendix~\ref{Sec:A}) in first-order perturbation theory (Hartree-Fock approximation) are
\begin{equation}
i\Sigma^{jj(1)}_{\alpha}(k,xt) \propto \tilde{g} \int \frac{dk^{\prime} d\varepsilon^{\prime}}{(2\pi)^2}(k-k^{\prime})G^{-+}_{\alpha}(k^{\prime} \varepsilon^{\prime},xt),
\end{equation}
where $j=-,+$. However, the off-diagonal elements $\Sigma^{-+(1)}_\alpha$ and $\Sigma^{+-(1)}_\alpha$ are equal to zero, so there is no entropy production in first-order perturbation theory.

\subsection{Two-particle collisions}
\label{Two-particle collision}
We now consider the self-energies $\Sigma^{ij}_{\alpha}$ up to the second order in the perturbation expansion,
\begin{equation}
\label{Self energy for right and left movers}
\begin{split}
& \Sigma^{ii^{\prime}(2)}_{\alpha}(k_1 \varepsilon_1; xt) \propto \int  \prod_{l=2,1',2'} \frac{dk_l d\varepsilon_l}{(2\pi)^2}\left| \mathcal{A}^{k^{\prime}_1, k^{\prime}_2}_{k_1,k_2}\right|^2 \delta(E-E^{\prime}) \\
& \times G^{ii^{\prime}}_{\alpha}(k^{\prime}_1 \varepsilon^{\prime}_1;xt) G^{i^{\prime}i}_{\tilde{\alpha}}(k_2 \varepsilon_2;xt) G^{ii^{\prime}}_{\tilde{\alpha}}(k^{\prime}_2 \varepsilon^{\prime}_2;xt),
\end{split}
\end{equation}
where $i,i^{\prime}=-,+$, $E=\varepsilon_1+\varepsilon_2$, $E^{\prime}=\varepsilon_{1^{\prime}}+\varepsilon_{2^{\prime}}$, $\varepsilon_l=k^2_l/2\tilde{m}+\alpha \tilde{v}_F k_l$, $\tilde{\alpha}=-\alpha$ and the scattering factor has the form
\begin{equation}
\left| \mathcal{A}^{k^{\prime}_1, k^{\prime}_2}_{k_1,k_2}\right|^2 =\tilde{g}^2 (k_1-k^{\prime}_2)^2 \delta(k_1+k_2-k^{\prime}_1-k^{\prime}_2).
\end{equation}
Furthermore, we simplify the quantum kinetic equation by using the quasiparticle approximation for the spectral function. In contrast to the interactions between the physical fermions, the quasiparticle interactions~(\ref{Interaction Hamiltonian of fermion quasiparticles}) are RG-irrelevant, so they cause only small corrections to the free spectral function. In quasiparticle approximation, the self-energy is dropped in the retarded Green's function $G^R_{\alpha}=(G^{-1}_{0,\alpha}-\Sigma^R_{\alpha})^{-1} \approx G^{R}_{0,\alpha}$. Consequently, using Eq.~(\ref{Zero order Green;s function in Wigner representation}), the spectral function of the quasiparticles, $A_{\alpha}=-2\text{Im} G^R_{\alpha}$, has the form of a Dirac delta function
\begin{equation}
A_{\alpha}(k\varepsilon,xt) \approx 2\pi \delta(\varepsilon-\xi_{k,\alpha}).
\end{equation}
Therefore, in the kinetic equation we neglect terms with $\Sigma^R_{\alpha}$ on its left-hand side and using Eq.~(\ref{Quantum kinetic equation}) the result is given by
\begin{equation}
\left\{G^{-1}_{0,\alpha},A_{\alpha} \phi_{\alpha}\right\} = \mathcal{I}_{\alpha}.
\end{equation}
Next, using the definition of Poisson brackets, Eq.~(\ref{Poisson brackets}), and performing an integration with respect to energy variable $\varepsilon$, the corresponding equations can be expressed in terms of distribution functions in phase space
\begin{equation}
\label{Distribution function in weak-coupling limit}
f_{\alpha}(k,xt) \equiv \int \frac{d\varepsilon}{2\pi}A_{\alpha}(k\varepsilon,xt)\phi_{\alpha}(k\varepsilon,xt) \approx \phi_{\alpha}(k\xi_{k_{\alpha}},xt).
\end{equation}
In terms of these, the kinetic equations in quasiparticle approximation take the form
\begin{equation}
\label{Kinetic equation in case of two-particle collisions within quasiparticle approximation}
\frac{\partial f_{\alpha}}{\partial t}+\frac{k+\alpha \tilde{m} \tilde{v}_F}{\tilde{m}}\frac{\partial f_{\alpha}}{\partial x}-\sqrt{K}\frac{\partial U}{\partial x} \frac{\partial f_{\alpha}}{\partial k} = I_{k,\alpha}[f_R,f_L],
\end{equation}
where the collision term for two-particle collisions is given by
\begin{equation}
\label{Scattering integral for two-particle collision case within quasiparticle approximation}
I_{k_1,\alpha}\propto  \int  \prod_{i=2,1^{\prime},2^{\prime}}  \frac{dk_i}{2\pi}\left|\mathcal{A}^{k^{\prime}_1,k^{\prime}_2}_{k_1,k_2} \right|^2 \delta(E-E^{\prime})\left[\mathcal{F}_1-\mathcal{F}_2\right],
\end{equation}
where $\mathcal{F}_1=(1-f_{1,\alpha})(1-f_{2,\tilde{\alpha}})f_{1^{\prime},\alpha}f_{2^{\prime},\tilde{\alpha}}$ and $\mathcal{F}_2=f_{1,\alpha}f_{2,\tilde{\alpha}}(1-f_{1^{\prime},\alpha})(1-f_{2^{\prime},\tilde{\alpha}})$ and $f_{i,\alpha} \equiv f_{\alpha}(k_i,xt)$.

The corresponding expression for the entropy production is obtained by performing an integration over $k$ in Eq.~(\ref{Scattering integral for two-particle collision case within quasiparticle approximation}). The result of this integration is
\begin{equation}
\label{Entropy production in quasiparticle approximation Two-particle collision}
\begin{split}
& \left[\frac{\partial \mathcal{S}}{\partial t}\right]_{coll} \propto \int\prod_{\substack{i=1,2,\\1',2'}}  \frac{dk_i}{2\pi} \delta(E-E^{\prime}) \\
&  \times \left|\mathcal{A}^{k^{\prime}_1,k^{\prime}_2}_{k_1,k_2}\right|^2 \left[\mathcal{F}_1-\mathcal{F}_2\right] \log\left[\mathcal{F}_1/\mathcal{F}_2\right].
\end{split}
\end{equation}
Because of $(x-y)\log(x/y) >0$ for positive $x$ and $y$, one finds indeed that $[\partial \mathcal{S}/\partial t]_{coll} \geq 0 $, in agreement with the second law of thermodynamics. However, due to the conservation laws of momentum $k_1+k_2=k^{\prime}_1+k^{\prime}_2$ and energy $\varepsilon_1+\varepsilon_2=\varepsilon^{\prime}_1+\varepsilon^{\prime}_2$, one actually finds that the inequality becomes an identity, and one recovers the same result as in equilibrium,
\begin{equation}
\left[\frac{\partial \mathcal{S}}{\partial t}\right]_{coll}=0.
\end{equation}
Thus, two-particle collisions alone cause no entropy production, and we have to consider the contribution due to three-particle collisions as the leading term. The vanishing effect of two-particle collisions due to kinematic constraints is already known for other relaxation phenomena in 1D systems.\cite{imambekov12}

\subsection{Three-particle collision}

In this section we investigate the effects of three-particle collisions and we will see that they provide the leading contribution to the entropy production. The left-hand side of the kinetic equation~(\ref{Kinetic equation in case of two-particle collisions within quasiparticle approximation}) retains the same form, but the scattering integral on the right-hand side now includes three-particle collisions. The scattering integral for this case has been already calculated using perturbation theory up to the fourth order.\cite{matveev13a,ristivojevic13,imambekov12,matveev19} In quasiparticle approximation, the resulting expression for the entropy production then has the form
\begin{equation}
\label{Entropy production in quasiparticle approximation three-particle collision}
\begin{split}
& \left[\frac{\partial \mathcal{S}}{\partial t}\right]_{coll} = \frac{1}{2}\int \prod_{\substack{l=1,2,3,\\1',2', 3'}} \frac{dk_l}{2\pi} \delta(E-E^{\prime}) \\
&  \times \left|\mathcal{A}^{k^{\prime}_1,k^{\prime}_2, k^{\prime}_3}_{k_1,k_2,k_3}\right|^2 \left[\mathcal{F}_1-\mathcal{F}_2\right] \log\left[\mathcal{F}_1/\mathcal{F}_2\right],
\end{split}
\end{equation}
where $E=\varepsilon_1+\varepsilon_2+\varepsilon_3$, $E^{\prime}=\varepsilon_{1^{\prime}}+\varepsilon_{2^{\prime}}+\varepsilon_{3^{\prime}}$. This expression includes conservation laws during the three-particle collision, and $|\mathcal{A}^{k^{\prime}_1,k^{\prime}_2, k^{\prime}_3}_{k_1,k_2,k_3}|^2$ is the square of the three-particle scattering amplitude between initial ($k_i$) and final states ($k^{\prime}_i$). The outgoing and incoming fluxes are given by $\mathcal{F}_1=(1-f_{1,R})(1-f_{2,R})(1-f_{3,L})f_{1^{\prime},R} f_{2^{\prime},R} f_{3^{\prime},L}$ and $\mathcal{F}_2=f_{1,R}f_{2,R}f_{3,L}(1-f_{1^{\prime},R}) (1-f_{2^{\prime},R})(1-f_{3^{\prime},L})$, respectively. The explicit form of scattering amplitude depends on the interaction Hamiltonian~(\ref{Interaction Hamiltonian of fermion quasiparticles in momentum representation}). In the case of the short-range interactions under consideration and at low temperatures $T/ \tilde{\varepsilon}_F \ll 1 $, the scattering amplitude squared takes the form~\citep{matveev13a,ristivojevic13,matveev19}
\begin{equation}
\left|\mathcal{A}^{k^{\prime}_1,k^{\prime}_2, k^{\prime}_3}_{k_1,k_2,k_3}\right|^2= \Lambda^2 (k_1-k_2)^2(k^{\prime}_1-k^{\prime}_2)^2\delta(\tilde{K}-\tilde{K}^{\prime}),
\end{equation}
where $\tilde{K}=k_1+k_2+k_3$ and $\tilde{K}^{\prime}=k_{1}^{\prime}+k_{2}^{\prime}+k_{3}^{\prime}$. For weakly interacting quasiparticles the constant prefactor $\Lambda$ is given in Refs.~[\onlinecite{matveev13a,ristivojevic13,matveev19}]. Noting that the inequality $(x-y)\log(x/y) \geq 0$ holds for any positive $x$ and $y$, we find again that the entropy production is compatible with the second law,
\begin{equation}
\left[\frac{\partial \mathcal{S}}{\partial t}\right]_{coll} \geq 0.
\end{equation}

To calculate the entropy production explicitly, we need to solve the quantum kinetic equation in quasiparticle approximation, see Eq.~(\ref{Kinetic equation in case of two-particle collisions within quasiparticle approximation}). The scattering integral is a nonlinear functional of the distribution functions, so an exact solution of the integro-differential equation is not possible. However, for weak interactions and slow external perturbation one can expand the distribution function around the equilibrium one in orders of interaction strength and external perturbation strength, i.e $f_{k,\alpha} \simeq f^{eq}_{k,\alpha}+\delta f_{k,\alpha}$, where $f^{eq}_{k,\alpha}=1/\{1+\exp[(k^2/2\tilde{m}+\alpha \tilde{v}_F k)/T]\}$. Substituting this expansion into Eq.~(\ref{Kinetic equation in case of two-particle collisions within quasiparticle approximation}), we obtain the following partial differential equation
\begin{equation}
\label{Partial differential equation for corrections to distribution function}
\frac{\partial}{\partial t} \delta f_{k,\alpha}+\frac{k+\alpha \tilde{m}\tilde{v}_F}{\tilde{m}}\frac{\partial}{\partial x}\delta f_{k,\alpha} +\sqrt{K}F(x,t)\frac{\partial f^{eq}_{k,\alpha}}{\partial k}=0,
\end{equation}
where we introduced the force $F(x,t)=-\partial U(x,t)/\partial x$. The solution of Eq.~(\ref{Partial differential equation for corrections to distribution function}) is given by
\begin{equation}
\label{Solution of kinetic equation}
\delta f_{k,\alpha}(x,t)=\zeta_{k,\alpha}(x,t)+\frac{\partial f^{eq}_{k,\alpha}}{\partial k} \int^t_{-\infty}dt^{\prime}\tilde{F}_{\alpha}(x,t^{\prime}),
\end{equation}
where $\tilde{F}_{\alpha}(x,t^{\prime})=\sqrt{K}F[x-(k+\alpha \tilde{m}\tilde{v}_F)(t-t^{\prime})/\tilde{m}, t^{\prime}]$ and $\zeta_{k,\alpha}(x,t)=\Phi_{\alpha}[x-(k+\alpha \tilde{m}\tilde{v}_F)t/\tilde{m}]$ is parametrized by an arbitrary function $\Phi_\alpha(x)$. The arbitrariness in $\Phi_\alpha(x)$ reflects the possibility to choose initial conditions. For further calculations we set $\zeta_{k,\alpha}(x,t)=0$ such that at $t \to -\infty$ we get the position-independent equilibrium distribution function.

Now we substitute the total distribution function, $f_{k,\alpha} \simeq f^{eq}_{k,\alpha}+\delta f_{k,\alpha}$ into Eq.~(\ref{Entropy production in quasiparticle approximation three-particle collision}) and expand in the correction. Due to conservation laws of momentum and energy, all non-vanishing corrections are second-order terms in $\delta f_{k,\alpha}$. Consequently, the expression for entropy production takes the following form
\begin{equation}
\label{Entropy production in three-particle case after the linearization of distribution function}
\begin{split}
& \left[\frac{\partial \mathcal{S}}{\partial t}\right]_{coll} =\int \prod_{\substack{l=1,2,3,\\1',2', 3'}} \frac{dk_i}{2\pi} \left|\mathcal{A}^{k^{\prime}_1,k^{\prime}_2,k^{\prime}_3}_{k_1,k_2,k_3}\right|^2 \times \mathcal{F}^{eq}_1 \times \delta(E-E^{\prime}) \\
& \times \left(\chi_{1,R}+\chi_{2,R}+\chi_{3,L}-\chi_{1^{\prime},R}-\chi_{2^{\prime},R}-\chi_{3^{\prime},L} \right)^2,
\end{split}
\end{equation}
where $\chi_{i,\alpha}=\delta f_{i,\alpha}/[f^{eq}_{i,\alpha}(1-f^{eq}_{i,\alpha})]$ and $\mathcal{F}^{eq}_1=(1-f^{eq}_{1,R})(1-f^{eq}_{2,R})(1-f^{eq}_{3,L})f^{eq}_{1^{\prime},R}f^{eq}_{2^{\prime},R}f^{eq}_{3^{\prime},L}$.
Substituting Eq.~(\ref{Solution of kinetic equation}) into Eq.~(\ref{Entropy production in three-particle case after the linearization of distribution function}) and taking into account the conservation laws for energy and momentum,
\begin{equation}
\begin{split}
& k'_{2}\simeq (k_1-k_{1^{\prime}}+k_2)+\frac{(k_1-k_{1}^{\prime})(k_{1}^{\prime}-k_2)}{2\tilde{m}\tilde{v}_F}+O\left[\frac{1}{\tilde{m}}\right]^2, \\
& k'_{3} \simeq k_3-\frac{(k_1-k_{1}^{\prime})(k_{1}^{\prime}-k_2)}{2\tilde{m}\tilde{v}_F}+O\left[\frac{1}{\tilde{m}}\right]^2, \\
\end{split}
\end{equation}
which we approximated to first order in band curvature, we obtain the final result for entropy production per unit time and unit length,
\begin{equation}
\label{Total entropy production. Final result}
\left[\frac{\partial \mathcal{S}}{\partial t}\right]_{coll} \simeq 16 \gamma \left(\Lambda \tilde{m}^2 \tilde{\varepsilon}_F\right)^2  \frac{g^2(x,t)}{\tilde{v}_F} \left(\frac{2T}{\tilde{\varepsilon}_F}\right)^{10},
\end{equation}
where $\gamma$ is a dimensionless prefactor of order one shown in App.~\ref{Sec:C} and
\begin{equation}\label{eq:g}
g(x,t)=\int \frac{d q}{2\pi} \int \frac{d\omega}{2\pi} \frac{\tilde{v}^2_F q^2 \omega \sqrt{K}U(q,\omega)}{(\omega-\tilde{v}_F q)^3}e^{iqx-i\omega t}
\end{equation}
encapsulates the dependence on the external perturbation. Integrating over the length of the 1D system, the entropy current term vanishes and one finds that the total change in entropy is given by the total entropy production rate,
\begin{align}
    \dot{S}
&=
    \frac{d}{dt} \int dx \sum_{\alpha = R,L} s_\alpha(x,t)
=
    \int dx \left[\frac{\partial \mathcal{S}}{\partial t}\right]_{coll}
=:
    \Delta \dot{S},
\end{align}
which is given by
\begin{align}\label{eq:DeltaS3}
\Delta \dot{S}(t)
&= 16 \gamma \left(\Lambda \tilde{m}^2 \tilde{\varepsilon}_F\right)^2 \left(\frac{2T}{\tilde{\varepsilon}_F}\right)^{10} \frac{1}{\tilde{v}_F} \int_0^L dx g^2(x,t).
\end{align}
This quantity is non-negative in accordance with the second law of thermodynamics. Moreover, it vanishes towards zero temperature as required by the third law. For a spatially homogeneous perturbation $U(x,t) \equiv U(t)$, one finds that $\Delta \dot{S} = 0$. This is reasonable because such a perturbation would correspond to a global variation of the chemical potential, which can be gauged away and is thus not expected to produce entropy. Moreover, using Eq.~(\ref{eq:g}), one finds that a time-independent perturbation $U(x,t) \equiv U(x)$ would lead to zero energy production as well. More precisely, the entropy production scales as $[\partial_t U(x,t)]^2$, the same scaling for slow drive as found in Refs.~[\onlinecite{bruch16,haughian18}] for a driven resonant level. The linear Luttinger liquid limit can be reached by taking the limit $\tilde{m} \to \infty$ at constant $\tilde{v}_F$, in which case one finds $\Delta \dot{S} = 0$ as expected.

Equation~(\ref{eq:DeltaS3}) represents the main result of our work. It applies to 1D quantum systems at arbitrary interaction strength and shows how entropy is produced by an external drive depending on both space and time. The entropy production can in principle be studied experimentally thanks to Eq.~(\ref{eq:DeltaS_WF}) by comparing the absorbed work with the change of free energy in the system.

\section{Conclusion}
\label{Sec:Conclusion}

In this paper, we have studied nonequilibrium thermodynamics of one-dimensional electron systems. Compared to their higher dimensional counterparts, one-dimensional systems pose the additional challenge that interactions cannot be investigated using perturbation theory, even if they are weak. Luttinger liquid theory provides a convenient framework for studying interacting 1D systems at low energy, but the linearization of the spectrum is too crude an approximation for studying relaxation or more general thermodynamic phenomena. This raises the question of how a nonequilibrium entropy, which must be consistent with the second law of thermodynamics, can be defined. To study this question, we considered a 1D electron system in a slowly varying external perturbation which brings the system out of equilibrium. We then used refermionization to express the system in terms of fermionic quasiparticles. In contrast to the physical electrons, the effective interactions between the quasiparticles are weak and allow a perturbative investigation.

Using the gradient approximation as well as perturbation theory, we derived a quantum kinetic equation which served as a basis for the definition of the full nonequilibrium entropy. We showed that this entropy satisfies a continuity equation whose source term is the entropy production. Kinematic constraints specific to one dimension mean that two-particle scattering does not lead to a nonzero entropy production. We therefore identified three-particle scattering as the leading process giving rise to a positive entropy production and determined its scaling at low temperature.

In equilibrium, our definition of the entropy coincides with an expression found previously for 1D systems by Rozhkov.\cite{rozhkov05,rozhkov06} For free fermions, the quasiparticles become identical to the physical fermions, and our entropy coincides with previously found expressions for the nonequilibrium entropy. \cite{bruch16,haughian18} For free particles at equilibrium, it coincides with the well-known von Neumann entropy of free fermions.

We expect our results to describe the nonequilibrium thermodynamics of 1D systems at low energies. Our treatment accounts for the band curvature and therefore allows us to exceed the ``zero-energy'' limit of (linear) Luttinger theory. However, towards higher temperatures, the quasiparticle interactions are known to become stronger and the bosonic basis becomes a more suitable starting point for a perturbative analysis.\cite{protopopov14} A similar analysis based on these bosonic modes would deserve future investigation.

\acknowledgments
The authors acknowledge financial support from the Fonds National de la Recherche Luxembourg under the grants ATTRACT~7556175 and INTER~11223315.

\appendix
\begin{widetext}
\section{Dyson equation}
\label{Sec:A}
The nonequilibrium Keldysh technique has been widely used to study dynamical effects in condensed matter physics.\cite{keldysh65,rammer,kita,arseev} In this section we give a brief overview of Keldysh formalism to the extent needed for our calculations. The main objective is the Green's function, which is defined as follows\cite{kita} (for $\alpha, \beta \in \{R, L\}$)
\begin{equation}
G_{\alpha \beta}(1^C,2^C) \equiv -i\langle \mathcal{T}_C\{\psi_{\mathcal{H},\alpha}(1^C)\psi^{\dagger}_{\mathcal{H},\beta}(2^C)\}\rangle=-i\langle \mathcal{T}_C \mathcal{S}_C\psi_{\alpha}(1^C)\psi^{\dagger}_{\beta}(2^C)\rangle_{0,{\rm conn}}.
\end{equation}
Here, the subscript $\mathcal{H}$ denotes time evolution in the Heisenberg picture, $\mathcal{T}_C$ is the time-ordering operator on the Keldysh contour, and the fermion quasiparticle operators $\psi_{\alpha}$ on right hand side are written in interaction picture with $\mathcal{H}_0(t)$ as unperturbed Hamiltonian. The notation $1^C \equiv (x_1,t^C_1)$ refers to spacetime coordinates where the time $t^C_1$ is on the Keldysh contour. The subscript ``$\rm conn$'' stands for connected diagrams. The average $\langle ...\rangle_0$ is taken with respect to the ground state of the Hamiltonian $\mathcal{H}_0(t_0 \to -\infty)$ of Eq.~(\ref{Total Hamiltonian of fermion quasiparticles}), at which time we assume that the system is in a thermal equilibrium state described by the grand canonical ensemble. The scattering matrix is given by
\begin{equation}
\label{Scattering matrix}
\mathcal{S}_C \equiv \mathcal{T}_C\exp\left[-i\sum_{\eta = \pm} (-\eta) \int ds^\eta \mathcal{H}_{int}(s^\eta)\right],
\end{equation}
where the two-body interaction Hamiltonian is presented in Eqs.~(\ref{Interaction Hamiltonian of fermion quasiparticles}) and (\ref{Interaction Hamiltonian of fermion quasiparticles in momentum representation}). Here, we have introduced the superscripts $\eta = \pm$ to distinguish the time variables on different Keldysh branches. Times on the forward branch (from $-\infty$ to $\infty$) are denoted by $t^-$, whereas $t^+$ refers to times on the backward branch ($\infty$ to $-\infty$). Using the definitions of the Green's function and scattering matrix, one can construct the perturbation theory.

For this purpose, we define the four Keldysh components of the Green's function, $G^{\eta\eta'}_{\alpha}(1,2)\equiv G_{\alpha\alpha}(1^\eta,2^{\eta'})$, and use them to construct the following matrix
\begin{equation}
\label{Keldysh matrix}
\check{G}_{\alpha}(1,2) = \begin{bmatrix}
       G^{--}_{\alpha}(1,2) & G^{-+}_{\alpha}(1,2)         \\[0.3em]
       G^{+-}_{\alpha}(1,2) & G^{++}_{\alpha}(1,2)            \\[0.3em]
     \end{bmatrix}.
\end{equation}
The exact expressions for the matrix elements are given by
\begin{equation}
\begin{split}
& G^{+-}_{\alpha}(1,2)
=-i\langle \psi_{\mathcal{H},\alpha}(1)\psi^{\dagger}_{\mathcal{H},\alpha}(2)\rangle, \\
& G^{-+}_{\alpha}(1,2)
=i\langle \psi^{\dagger}_{\mathcal{H},\alpha}(2)\psi^{\dagger}_{\mathcal{H},\alpha}(1)\rangle, \\
& G^{--}_{\alpha}(1,2)=\theta(t_1-t_2)G^{+-}_{\alpha}(1,2)+\theta(t_2-t_1)G^{-+}_{\alpha}(1,2), \\
& G^{++}_{\alpha}(1,2)=\theta(t_2-t_1)G^{+-}_{\alpha}(1,2)+\theta(t_1-t_2)G^{-+}_{\alpha}(1,2),
\end{split}
\end{equation}
where $\theta(t)$ is the Heaviside step function. The self-energy matrix has a similar form, namely
\begin{equation}
\label{Self energy matrix}
\check{\Sigma}_{\alpha}(1,2) = \begin{bmatrix}
       \Sigma^{--}_{\alpha}(1,2) & \Sigma^{-+}_{\alpha}(1,2)         \\[0.3em]
       \Sigma^{+-}_{\alpha}(1,2) & \Sigma^{++}_{\alpha}(1,2)            \\[0.3em]
     \end{bmatrix}.
\end{equation}
Using the above notations, we can express the Dyson equation in differential and integral forms as follows~\cite{kita}
\begin{equation}
\label{Differential and integral Dyson equations}
\begin{split}
& \left(i\partial_{t_1}-\mathcal{K}_{\alpha,1}\right)\check{G}_{\alpha}(1,2)-\int d3\ \check{\Sigma}_{\alpha}(1,3)\check{\sigma}_3\check{G}_{\alpha}(3,2)=\check{\sigma}_3\delta(1,2), \\
& \check{G}_{\alpha}(1,2)=\check{G}^{(0)}_{\alpha}(1,2)+\int d3\ \int d4\ \check{G}^{(0)}_{\alpha}(1,3)\check{\sigma}_3 \check{\Sigma}_{\alpha}(3,4)\check{\sigma}_3\check{G}_{\alpha}(4,2),
\end{split}
\end{equation}
where $\check{\sigma}_3$ is the Pauli matrix and $\delta(1,2)=\delta(t_1-t_2)\delta(x_1-x_2)$. It is worth noting that $G^{(0)}_{\alpha \beta}(1,2)=\delta_{\alpha \beta} G^{(0)}_{\alpha}(1,2)$. Thus initially right (R) and left (L) fermionic quasiparticle fields do not correlate, for instance $G^{(0)}_{L R}(1,2)=0$.

For mathematical convenience it is useful to introduce Green's functions in a rotated basis used by Larkin and Ovchinnikov~\cite{larkin}
\begin{equation}
\label{Keldysh matrix after rotation}
\check{G}^K_{\alpha}(1,2) := \check{L}\check{\sigma}_3 \check{G}_{\alpha}\check{L}^{\dagger}=\begin{bmatrix}
       G^{R}_{\alpha} & G^{K}_{\alpha}         \\[0.3em]
       0 & G^{A}_{\alpha}           \\[0.3em]
     \end{bmatrix},
     \quad \text{where} \quad
     \check{L}=\frac{1}{\sqrt{2}}
     \begin{bmatrix}
       1 & -1         \\[0.3em]
       1 & 1           \\[0.3em]
     \end{bmatrix}.
\end{equation}
The matrix elements of $\check{G}^K_{\alpha}(1,2)$ are the well-known Keldysh, retarded and advanced Green's functions. They can be rewritten as $G^R_{\alpha}(1,2)=\theta(t_1-t_2)[G^{+-}_{\alpha}(1,2)-G^{-+}_{\alpha}(1,2)]$, $G^A_{\alpha}(1,2)=-\theta(t_2-t_1)[G^{+-}_{\alpha}(1,2)-G^{-+}_{\alpha}(1,2)]$ and $G^K_{\alpha}(1,2)=G^{+-}_{\alpha}(1,2)+G^{-+}_{\alpha}(1,2)$. The self-energy matrix~(\ref{Self energy matrix}) can be converted to a form similar to Eq.~(\ref{Keldysh matrix after rotation}). Using the Larkin-Ovchinnikov basis we can write the Dyson equation for $\check{G}^K_{\alpha}$ in the form
\begin{equation}
\left(i\partial_{t_1} - \mathcal{K}_{\alpha,1}\right)\check{G}^K_{\alpha}(1,2)-\int d3\ \check{\Sigma}^K_{\alpha}(1,3)\check{G}^K_{\alpha}(3,2)=\check{I}\delta(1,2),
\end{equation}
where $\check{I}$ is a $2\times 2$ identity matrix.

\section{Derivation of continuity equation}
\label{Sec:B}

Let us multiply the quantum kinetic equation in Eq.~(\ref{Quantum kinetic equation}) by $\log\left[(1-\phi_{\alpha})/\phi_{\alpha}\right]$, carry out the integrations over $k$ and $\varepsilon$, and make use of $\text{Im}\{(G^R_{\alpha})^{-1}, G^R\}=\text{Im}\{G^{-1}_{0,\alpha}-\Sigma^R_{\alpha},G^R_{\alpha}\}=0$ as well as $\log\left[(1-\phi_{\alpha})/\phi_{\alpha}\right]d\phi_{\alpha}=d\sigma_{\alpha}$, where the function $\sigma_{\alpha}[\phi_{\alpha}]$ is given by Eq.~(\ref{General expression for entropy}) of main text. Straightforwardly, we get the following expression from kinetic equation
\begin{equation}
\label{Kinetic equation multiplied by log}
\int \frac{dk d\varepsilon}{(2\pi)^2}\left[\{G^{-1}_{0,\alpha}- \text{Re}\Sigma^R_{\alpha}, A_{\alpha}\phi_{\alpha}\}-\{A_{\Sigma_{\alpha}}\phi_{\alpha}, \text{Re}G^{R}_{\alpha}\}\right]\log\left(\frac{1-\phi_{\alpha}}{\phi_{\alpha}}\right)=\int \frac{dk d\varepsilon}{(2\pi)^2} \mathcal{I}_{\alpha}\log\left(\frac{1-\phi_{\alpha}}{\phi_{\alpha}}\right),
\end{equation}
where we introduced the notation $A_{\Sigma_{\alpha}}=-i\phi^{-1}_{\alpha} \Sigma^{-+}_{\alpha}$. According to Eq.~(\ref{Entropy production for right and left mover}) the right-hand side is  the entropy production per unit time and unit length. To simplify the integrand of left-hand side of Eq.~(\ref{Kinetic equation multiplied by log}), we use the following property of Poisson brackets
\begin{equation}
\{b,a\phi_{\alpha}\}\log \phi_{\alpha}+\{b,a(1-\phi_{\alpha})\}\log(1- \phi_{\alpha})=\{b,a[\phi_{\alpha} \log \phi_{\alpha}+(1-\phi_{\alpha})\log (1-\phi_{\alpha})]\}=-\{b,a\sigma_{\alpha}[\phi_{\alpha}]\}.
\end{equation}
This expression holds for arbitrary functions $a$, $b$ and positive $0<\phi_{\alpha}<1$. Using this identity we rewrite the integrand (a constant prefactor $1/(2\pi)^2$ is omitted) of left-hand side of Eq.~(\ref{Kinetic equation multiplied by log}) in the form
\begin{equation}
\begin{split}
& \left[\{G^{-1}_{0,\alpha}- \text{Re}\Sigma^R_{\alpha}, A_{\alpha}\phi_{\alpha}\}-\{A_{\Sigma_{\alpha}}\phi_{\alpha}, \text{Re}G^{R}_{\alpha}\}\right]\log\left(\frac{1-\phi_{\alpha}}{\phi_{\alpha}}\right)=\{G^{-1}_{0,\alpha}- \text{Re}\Sigma^R_{\alpha}, A_{\alpha}\sigma_{\alpha}[\phi_{\alpha}]\}+\{ \text{Re}G^{R}_{\alpha},A_{\Sigma_{\alpha}}\sigma_{\alpha}[\phi_{\alpha}]\}+\\
& + \left[\{G^{-1}_{0,\alpha}-\text{Re}\Sigma^R_{\alpha},A_{\alpha}\}+\{\text{Re}G^R_{\alpha},A_{\Sigma_{\alpha}}\}\right]\log(1- \phi_{\alpha}).
\end{split}
\end{equation}
Next, using the relation $\text{Im}\{G^{-1}_{0,\alpha}-\Sigma^R_{\alpha},G^R_{\alpha}\}=0$ one can show that the last term of the above expression is equal to zero, i.e., $[\{G^{-1}_{0,\alpha}-\text{Re}\Sigma^R_{\alpha},A_{\alpha}\}+\{\text{Re}G^R_{\alpha},A_{\Sigma_{\alpha}}\}]\log(1-\phi_{\alpha})=0$.
Consequently, using the definition of the Poisson bracket in Eq.~(\ref{Poisson brackets}) we obtain that the left-hand side of Eq.~(\ref{Kinetic equation multiplied by log}) takes the form
\begin{align}
\begin{split}
&\int \frac{dk d\varepsilon}{(2\pi)^2}\left[\{G^{-1}_{0,\alpha}- \text{Re}\Sigma^R_{\alpha}, A_{\alpha}\sigma_{\alpha}[\phi_{\alpha}]\}+\{ \text{Re}G^{R}_{\alpha},A_{\Sigma_{\alpha}}\sigma_{\alpha}[\phi_{\alpha}]\}\right]=\frac{\partial}{\partial t} \underbrace{\int \frac{dk d\varepsilon}{(2\pi)^2}\sigma_{\alpha} \left[A_{\alpha}\frac{\partial \left(G^{-1}_{0,\alpha}-\text{Re}\Sigma^R_{\alpha}\right)}{\partial \varepsilon}+A_{\Sigma_{\alpha}}\frac{\partial \text{Re}G^R_{\alpha}}{\partial \varepsilon}\right]}_{s^{(\alpha)}} \\
&-\left.\sigma_{\alpha} A_{\alpha} \frac{\partial \left(G^{-1}_{0,\alpha}-\text{Re}\Sigma^R_{\alpha}\right)}{\partial t} \right |_{\varepsilon \to -\infty}^{\varepsilon \to +\infty} -\left.\sigma_{\alpha} A_{\Sigma_{\alpha}} \frac{\partial \left(\text{Re}G^R_{\alpha}\right)}{\partial t} \right |_{\varepsilon \to -\infty}^{\varepsilon \to +\infty}+\frac{\partial}{\partial x} \underbrace{\int \frac{dk d\varepsilon}{(2\pi)^2}\sigma_{\alpha} \left[-A_{\alpha}\frac{\partial \left(G^{-1}_{0,\alpha}-\text{Re}\Sigma^R_{\alpha}\right)}{\partial k}-A_{\Sigma_{\alpha}}\frac{\partial \text{Re}G^R_{\alpha}}{\partial k}\right]}_{j^{(\alpha)}_s}\\
& + \left.\sigma A_{\alpha} \frac{\partial \left(G^{-1}_{0,\alpha}-\text{Re}\Sigma^R_{\alpha} \right)}{\partial x} \right |_{k \to -\infty}^{k \to +\infty}+\left.\sigma_{\alpha} A_{\alpha} \frac{\partial \left(\text{Re}G^R_{\alpha}\right)}{\partial x} \right |_{k \to -\infty}^{k \to +\infty}.
\end{split}
\end{align}
Omitting the boundary terms we arrive at Eq.~(\ref{Continuity equation}) of main text.

\section{Value of the dimensionless prefactor}
\label{Sec:C}
\begin{equation}
\begin{split}
& \gamma = \frac{1}{4^{10}} \int dx_1  \int dx_2 \int dx^{\prime}_1 \int dx_3 (x_1-x^{\prime}_1)^2 (x_2-x^{\prime}_1)^2 (x_1-x_2)^2(2x^{\prime}_1-x_1-x_2)^2 \times \\
& \times \left(1-\frac{1}{1+e^{x_1}}\right) \left(1-\frac{1}{1+e^{x_2}}\right) \frac{1}{1+e^{x^{\prime}_1}}
\frac{1}{1+e^{x_1+x_2-x^{\prime}_1}} \left(1-\frac{1}{1+e^{-x_3}}\right)\frac{1}{1+e^{-x_3}} \approx 0.8823.
\end{split}
\end{equation}
\end{widetext}

\bibliography{refsclean}

\end{document}